\documentclass[useAMS,usenatbib]{mn2e}
\usepackage{graphicx}
\usepackage{amsmath}
\usepackage{amssymb}
\usepackage{color}
\usepackage{url}

\newcommand{\wdth}{SDSS 1355$+$0856}

\newcommand{\kms}{km s$^{-1}$}
\newcommand{\msun}{M$_{\odot}$}

\title[\wdth]{\wdth: A detached white dwarf + M star binary in the period gap discovered by the SWARMS survey \thanks{Observations
    reported here were obtained at the MMT Observatory, a joint facility of the Smithsonian Institution and the University of
    Arizona}.}  \author[C. Badenes et al.]  {\parbox{\textwidth}{Carles Badenes$^{1}$\thanks{E-mail: badenes@pitt.edu} Marten
    H. van Kerkwijk$^{2}$, Mukremin Kilic$^{3}$, Steven J. Bickerton$^{4}$, Tsevi Mazeh$^{5}$, Fergal Mullally$^{6}$, Lev
    Tal-Or$^{5}$, Susan E. Thompson$^{6}$ }
  \vspace{0.4cm}\\
\parbox{\textwidth}{$^{1}$ Department of Physics and Astronomy and Pittsburgh Particle Physics, Astrophysics and Cosmology Center
  (PITT PACC), University of Pittsburgh, 3941 O'Hara St, Pittsburgh PA 15260, USA \\
  $^{2}$ Department of Astronomy and Astrophysics, University of Toronto, 50 St. George Street, Toronto, ON M5S 3H4, Canada. \\
  $^{3}$ Homer L. Dodge Department of Physics and Astronomy, University of Oklahoma, 440 W. Brooks St., Norman, OK, 73019, USA \\
  $^{4}$ Institute for the Physics of Mathematics of the Universe (IPMU), The University of Tokyo, Chiba 277-8582, Japan \\
  $^{5}$ School of Physics and Astronomy, Tel-Aviv University, Tel-Aviv 69978, Israel\\
  $^{6}$ SETI Institute/NASA Ames Research Center, Moffett Field, CA 94035, USA} \\
}

\begin{document}

\date{Accepted 0000. Received 0000; in original form 0000}


\maketitle

\label{firstpage}

\begin{abstract}
  \wdth\ was identified as a hot white dwarf (WD) with a binary companion from time-resolved SDSS spectroscopy as part of the
  ongoing SWARMS survey. Follow-up observations with the ARC 3.5m telescope and the MMT revealed weak emission lines in the
  central cores of the Balmer absorption lines during some phases of the orbit, but no line emission during other phases. This can
  be explained if \wdth\ is a detached WD+M dwarf binary similar to GD 448, where one of the hemispheres of the low-mass companion
  is irradiated by the proximity of the hot white dwarf. Based on the available data, we derive a period of $0.11438 \pm 0.00006$
  days, a primary mass of $0.46\pm0.01$ \msun, a secondary mass between $0.083$ and $0.097$ \msun, and an inclination larger than
  57$^{\circ}$. This makes \wdth\ one of the shortest period post-common envelope WD+M dwarf binaries known, and one of only a few where the
  primary is likely a He-core white dwarf, which has interesting implications for our understanding of common envelope
  evolution and the phenomenology of cataclysmic variables. The short cooling time of the WD ($25$ Myr) implies that the system
  emerged from the common envelope phase with a period very similar to what we observe today, and was born in the period gap
  of cataclysmic variables.
\end{abstract}

\begin{keywords}
binaries: close -- binaries: spectroscopic - stars: individual: \wdth -- white dwarfs.
\end{keywords}

\section{Introduction}

\noindent
The common envelope (CE) phase remains one of the key open issues in stellar evolution. Several important classes of astrophysical
sources go through at least one CE phase at some point in their lifetime, including cataclysmic variables (CVs), low mass X-ray
binaries, detached binary white dwarfs (WDs), AM CVn stars, and very likely the not-yet-identified progenitors of Type Ia
supernovae (SN Ia). During the CE phase, the two components of a binary system come into contact and create a shared atmosphere
that is ejected through friction, leading to a loss of energy and a drastic reduction of the orbital period
\citep{PaczynskiB.1976}. Due to the challenges involved in performing accurate numerical simulations of the CE phase
\citep{Taam2000a,Ricker2012}, theoretical studies have been largely restricted to simplified analytic calculations. One prescription that has
gained wide acceptance is the so-called $\alpha$ formalism, where the eponymous parameter represents the fraction of the orbital
energy loss that is consumed in unbinding the common envelope \citep[see][for a review]{Webbink2006}, but even the
 fundamental aspects of this prescription are still being revised \citep[see][]{Ivanova2011a,DeMarco2011,Zorotovic2011a,Davis2012}.

\begin{figure*}
  \includegraphics[width=0.32\textwidth,angle=90]{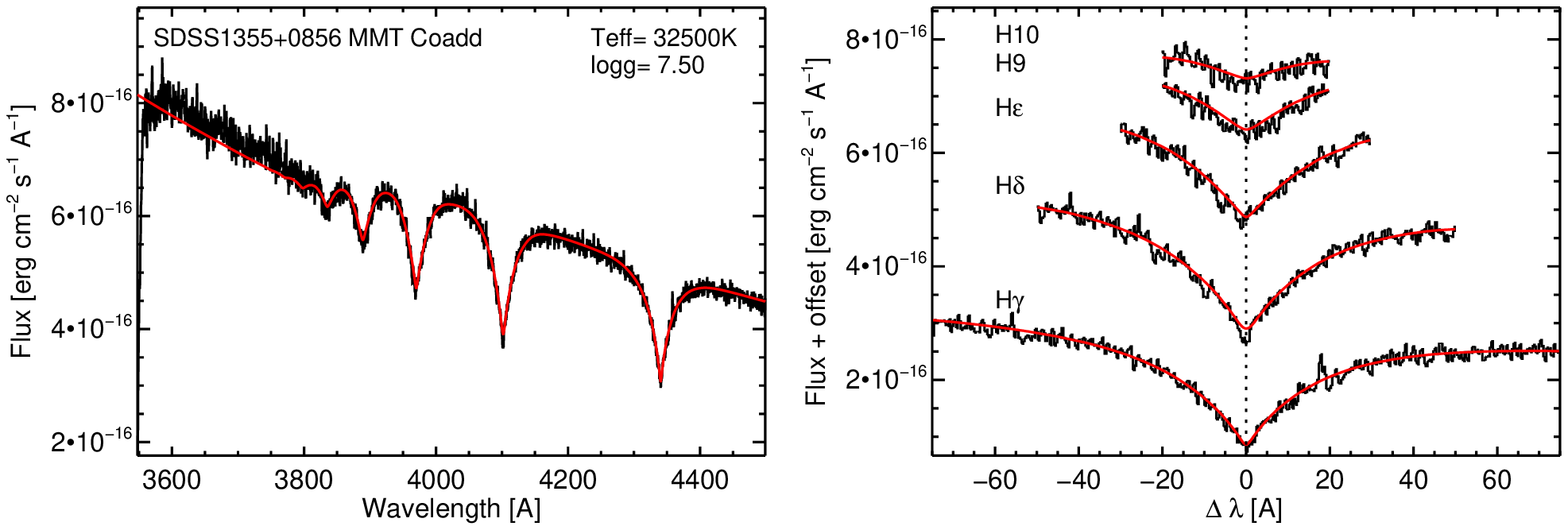}
  \caption{Co-added MMT spectrum of \wdth, together with the best-fit atmosphere model ($T_{\rm eff}=32500$ K, $\log{g}=7.50$). }
  \label{MMTspect}
\end{figure*}

The most useful constraints on CE evolution come from statistical studies of the properties of post-CE binaries: periods, mass
ratios, and demographics. \cite{Zorotovic2010} compiled a sample of all post-CE WD+main sequence star binaries (WDMS) known at the
time: 35 from the SDSS catalogue of \cite{rebassa10:SDSS_WD+M_catalog} plus 25 from the literature, with periods ranging between
$0.08$ and $21.72$ days. \citeauthor{Zorotovic2010} found that the properties of these 60 binaries can be explained by the
classical $\alpha$ formalism, with $\alpha$ between $0.2$ and $0.3$ \citep[but see][for a different analysis based on a smaller
sample of systems]{DeMarco2011}. In this context, post-CE systems with exceptionally high or low values of the period and/or the
masses of the components are especially interesting, because they can test the limits of the CE prescription and put important
constraints on the initial and final conditions of CE evolution. Post-CE binaries containing a He core WD ($M \la 0.48$ \msun;
\citealt{sweigart1978}) are of particular interest, because they anchor the low values of $\alpha$ \citep{Zorotovic2011}, and they
have shorter periods and less massive secondaries than their C/O core WD counterparts \citep{Zorotovic2011a}. The final fate of
these systems is also puzzling, because few, if any, CVs are known to have a He core WD primary, suggesting they might evolve into
classical novae with exceptionally rare outbursts \citep{Shen2009}.


In this paper, we report on SDSS J$135523.92+085645.4$ (henceforth \wdth), a short-period WD binary discovered by the SWARMS
survey. SWARMS \citep{badenes09:SWARMS_I,mullally09:DDWDs,Badenes2012,Maoz2012} exploits the time-resolved dimension in the
spectroscopic database of the Sloan Digital Sky Survey (SDSS) to identify short period double WD binaries \citep[see][for a brief
summary of the time resolved spectroscopy capabilities in SDSS]{Bickerton2012}. \wdth\ was originally targeted for follow-up
because it fulfilled the SWARMS selection criteria: evidence for radial velocity (RV) shifts between the sub-exposures in the SDSS
data and absence of features in the red part of the spectrum that might be associated with a non-degenerate companion (see Section
\ref{sec:Obs}). Follow-up observations with the ARC 3.5m telescope and the MMT, however, revealed the presence of weak emission
peaks in the Balmer cores of the spectral primary, indicating that the companion is not another WD, but a nondegenerate
object. The fact that these emission features are only present at certain orbital phases can be understood if they originate in
the irradiated side of the nondegenerate companion, a behaviour that has been previously noted in systems like GD 448 \citep[also
known as WD 0710+741 and LP 034-185][]{marsh1996MNRAS.278..565M:GD448,maxted1998MNRAS.300.1225M:GD448}. From our follow-up
observations (Section \ref{sec:Analysis}), we derive a period of $0.11438$ days for \wdth, placing it in the `period gap' seen in
the distribution of CVs, and making it one of the shortest-period post-CE binaries known. In Section \ref{sec:Disc}, we derive
estimates for the masses of the components and the orbital inclination, and we discuss the origin and final fate of the system.

\section{Observations}
\label{sec:Obs}

\subsection{SDSS and selection as a SWARMS candidate}

In the DR7 SDSS WD catalogue \citep{Kleinman2009}, \wdth\ is classified as an isolated DA WD with $T_{\rm eff}=33158 \pm 175$ K and
$\log{g}=7.37 \pm 0.04$.
To verify the basic properties of the WD, we performed an independent fit to a high S/N co-added spectrum obtained after removing
RV shifts from the MMT exposures without detected line emission (see Table \ref{tab:RVs} and Section \ref{sec:Analysis} for
details) using the latest version of the DA WD atmosphere models by Detlev Koester
\citep{finley97:WD_models_spectral,koester2010MmSAI..81..921K:WD_models}. Our fitted parameters are close to those of Kleinman et
al.: $T_{\rm eff}=32500 \pm 250$ K and $\log{g}=7.44 \pm 0.05$ (see Figure \ref{MMTspect}). The system was identified as a
candidate short-period binary by the SWARMS survey from the large RV shifts (roughly a hundred $\mathrm{km \, s^{-1}}$) found
between the four SDSS sub-exposures, which were taken with a temporal baseline of three days.

\begin{figure}
  \includegraphics[width=0.32\textwidth,angle=90]{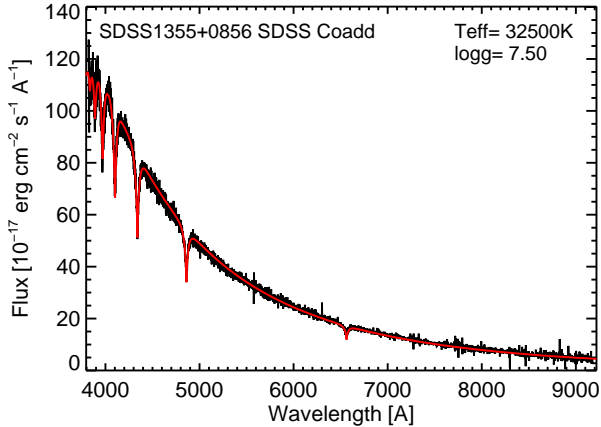}
  \caption{SDSS spectrum of \wdth, together with the best-fit atmosphere model ($T_{\rm eff}=32500$ K, $\log{g}=7.50$).}
  \label{fig:SDSSSpect}
\end{figure}

\begin{figure}
  \includegraphics[width=0.32\textwidth,angle=90]{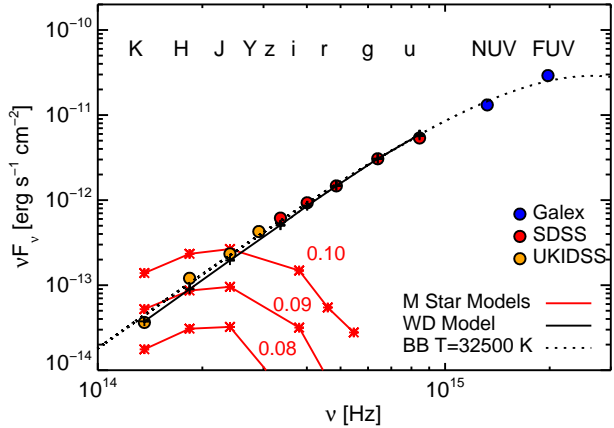}
  \caption{Spectral energy distribution for \wdth. The dotted black line is a blackbody spectrum with $T=32500$ K, normalized to
    match the observed flux in the $g$ band. The solid black line is a WD model from Holberg et al. (2006), with $T_{\rm eff}=35000$ K,
    $\log g=7.5$, and the same normalization. The solid red lines represent low-mass star models from Baraffe et al. (1998), labeled
    by mass in \msun.}
  \label{fig:SED}
\end{figure}

Because SWARMS often detects significant RV shifts from known WDMS binaries, we routinely cross-check all our candidate binaries
against the SDSS WDMS catalogue\footnote{Avialable on-line at \url{http://sdss-wdms.org}} of \cite{Rebassa-Mansergas2012} and
remove any matches from our follow-up schedule to maximize the efficiency in our discovery of short-period double WD
systems. \wdth, however does not appear in the SDSS WDMS catalogue. \cite{rebassa10:SDSS_WD+M_catalog} describe the procedure they
follow to identify WDMS binaries among the SDSS spectra. A key ingredient is the fitting of all candidate
objects with WD atmosphere models: any spectra that are well-fit by a WD model in the blue but not in the red are potential WDMS
binaries. \cite{rebassa10:SDSS_WD+M_catalog} calculate the $\chi^2$ in the red (7000 to 9000 \AA) and in the blue (4000 to 7000
\AA), and flag any spectra with $\chi^{2}_{r}/\chi^{2}_{b} > 1.5$ as WDMS binary candidates, which are then confirmed based on
their infrared photometry. As seen in Figure~\ref{fig:SDSSSpect}, a single-component WD model provides an excellent fit to the
SDSS spectrum of \wdth\ in the red - in fact, $\chi^{2}_{r}/\chi^{2}_{b} = 0.74$. This explains why the system is not listed in
the SDSS WDMS catalogue. For obvious reasons, it is hard to constrain the presence of low-mass M-type companions with hot WD
primaries; see Figure 9 of \cite{rebassa10:SDSS_WD+M_catalog} and the accompanying discussion for more details.

In addition to the SDSS broadband photometry, \wdth\ has NUV and FUV fluxes from \textit{Galex} \citep{Morrissey2007} and $Y$,
$J$, $H$, and $K$ IR photometry from UKIDSS \citep[][see Table~\ref{tab:Parameters}]{Lawrence2007}. The broadband SED is
reasonably well fit by a blackbody spectrum with $T=32500$ K, although the measured fluxes in the red and near IR present small
deviations with respect to the best isolated WD model in the grid of \cite{Holberg2006} ($T_{\rm eff}=35000$ K, $\log{g}=7.5$, see
Figure~\ref{fig:SED}).  When the WD model is normalized to match the measured $g$ flux, the SED of \wdth\ is brighter than the
model by $\sim$ 0.15 mag in $z$ and $J$, and $\sim$ 0.3 mag in $H$. It is hard to explain this flux excess as the contribution of
a low-mass stellar companion, because the WD model does match the flux in the $K$ band. A possible explanation for the excess is
that the IR data points were taken at different orbital phases, and that the excess in $J$ and $H$ comes from the irradiated side
of the companion, but given the known issues with the accuracy of UKIDSS photometry for objects fainter than $\sim18$ mag
\citep[see e.g.][]{Leggett2011}, the excess might also be purely instrumental.

\begin{table}
  \centering
  \caption{Observational Properties of \wdth}
  \begin{tabular}{@{\extracolsep{\fill}}lr}
    \hline
    Parameter & Value\\
    \hline
    R.A. (J2000) & $13$h $55$m $23.92$s\ \\
    Decl. (J2000) & $+8^\circ$ $56 ^\prime$ $45.4 ^{\prime \prime}$\\
    \hline
    \multicolumn{2}{c}{SED (GALEX/SDSS/UKIDSS)} \\
    \hline
    $FUV$ & $1487 \pm 60 \, \mathrm{\mu Jy}$\\
    $NUV$ & $971 \pm 27 \, \mathrm{\mu Jy}$\\
    $u$ & $17.05 \pm 0.01$ mag\\
    $g$ & $17.29 \pm 0.01$ mag\\
    $r$ & $17.76 \pm  0.02$ mag\\
    $i$ & $18.03 \pm 0.02$ mag\\
    $z$ & $18.26 \pm 0.03$ mag\\
    $Y$ & $17.85 \pm 0.02$ mag\\
    $J$ & $17.99 \pm 0.04$ mag\\
    $H$ & $17.98 \pm 0.07$ mag\\
    $K$ & $18.43 \pm 0.15$ mag\\
    \hline
    \multicolumn{2}{c}{Spectral Parameters (MMT)} \\
    \hline
    $T_{\rm eff}$ & $32050 \pm 350$ K \\
    $\log{g}$ & $7.44 \pm 0.05$ \\
    \hline
    \multicolumn{2}{c}{Orbital Parameters (MMT)} \\
    \hline
    $P$ & $0.11438 \pm 0.00006$ days \\
    $T_{0}$ & $55275.9613 \pm 0.0012$ MJD (bar) \\
    $K_{a}$ & $64 \pm 8 \, \mathrm{km \, s^{-1}}$ \\
    $\gamma_{a}$ & $-23 \, \pm 5 \mathrm{km \, s^{-1}}$ \\
    $K_{e}$ & $-261 \pm 13  \, \mathrm{km \, s^{-1}}$ \\
     $\Delta \gamma_{e-a}$ & $-4 \, \pm 18 \mathrm{km \, s^{-1}}$ \\
    \hline
    \multicolumn{2}{c}{Derived Parameters} \\
    \hline
    $M_{1}$ & $0.46 \pm 0.01$ \msun \\
    $M_{2}$ & $0.083 \leq M_{2} \leq 0.097$ \msun \\
    $i$ & $\geq 57^{\circ}$ \\
    $t_{Cool,WD}$ & $25$ Myr \\
    \hline
  \end{tabular}
  \label{tab:Parameters}
\end{table}

\subsection{Spectroscopic Follow-Up}

\begin{figure*}
  \includegraphics[width=0.9\textwidth]{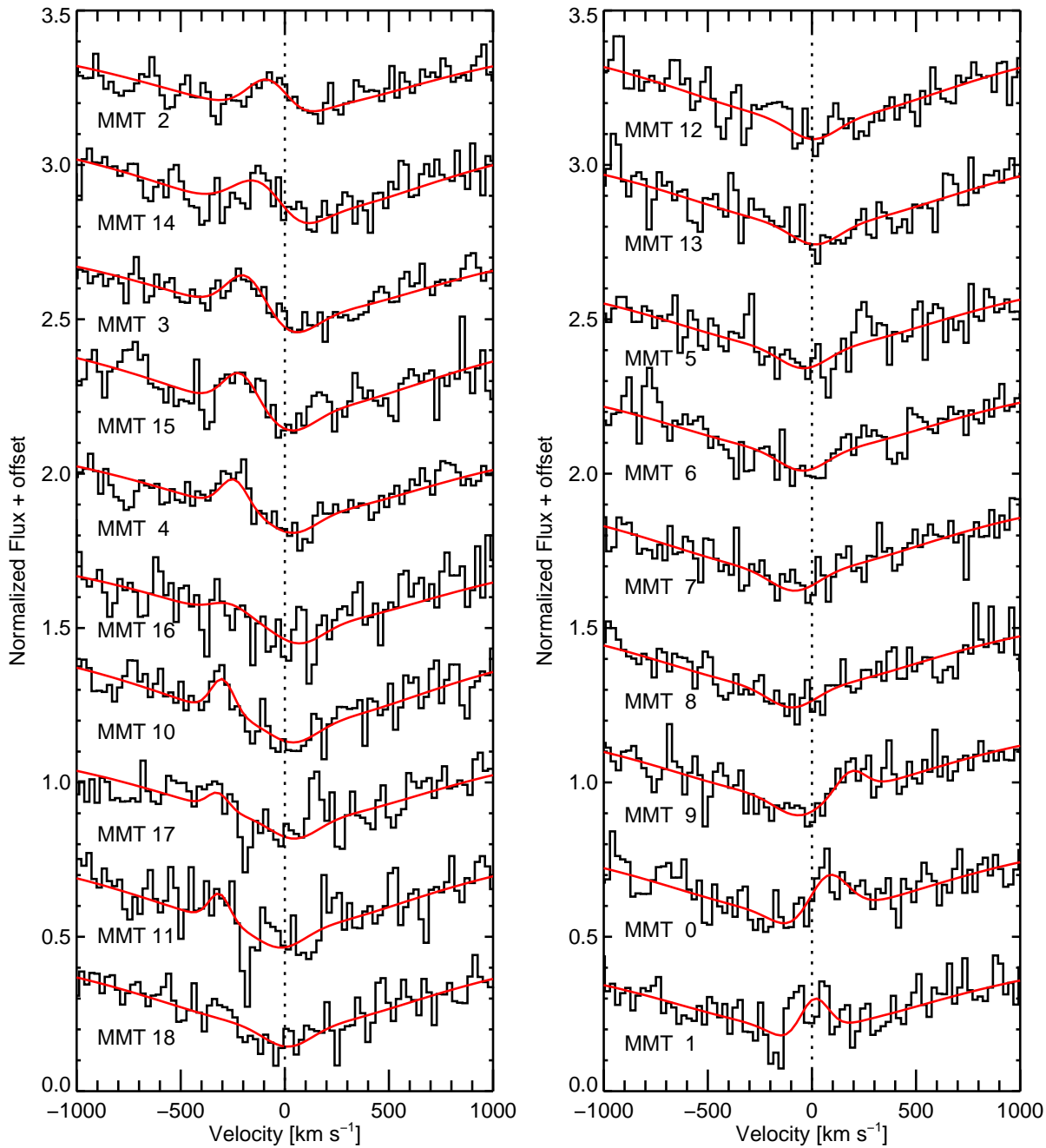}
  \caption{MMT spectra of \wdth\ around the H$\gamma$ line, ordered by phase (top to bottom and left to right). No line emission
    was detected in spectra MMT 18, 12, 13, 5, 6, 7, and 8. The red line is our best-fir model (see  Section \ref{sec:Analysis}
    for details).}
  \label{Spectra}
\end{figure*}

We observed \wdth\ with the Dual Imaging Spectrograph on the 3.5m Astrophysical Research Consortium telescope at Apache Point
Observatory on February 9 and 14, 2010. Variable cloud coverage and pointing issues due to strong winds resulted in extremely
noisy spectra. Clear RV shifts were apparent in some spectra, but in other cases the line cores had strange shapes, probably
because of inadequate pointing, and the RVs could not be determined with confidence. Because of these issues, we revisited \wdth\
with the 6.5m Multiple Mirror Telescope (MMT) at Mt. Hopkins observatory. We used the Blue Channel Spectrograph to obtain moderate
resolution spectroscopy of \wdth\ on March 19 and 21, 2010.  We operated the spectrograph with the 832 line mm$^{-1}$ grating in
second order and a 1$\arcsec$ slit, providing wavelength coverage 3600 \AA\ to 4500 \AA\ and a spectral resolution of 1.0 \AA.  We
obtained all observations at the parallactic angle, with a comparison lamp exposure paired with every observation.  We
flux-calibrated using blue spectrophotometric standards \citep{Massey1988}. Our observing and reduction procedures were similar to
the ones described in \cite{kilic10:WD_binaries_merge}. The superior quality of the MMT data revealed weak but clear emission
lines in some of the spectra, but no evidence of line emission in others (see Figure~\ref{Spectra}).

\section{Analysis}
\label{sec:Analysis}

The behaviour of the line emission revealed by the MMT data and shown on Figure \ref{Spectra} implies that the companion of \wdth\
cannot be a degenerate object, and is probably a faint main sequence star or a brown dwarf. The absence of line emission in some
of the spectra can be understood if the system is a binary similar to GD 448 \citep{marsh1996MNRAS.278..565M:GD448}, where one
hemisphere of the low-mass main sequence companion is irradiated by the proximity of the hot WD. This results in line emission
only during the phases in which most of the irradiated side of the secondary is visible, i.e. when the RV of the primary 
is increasing between its minimum and its maximum values. In this scenario, the secondary would not contribute
much flux to the red part of the SDSS spectrum (Figure ~\ref{fig:SDSSSpect}), especially if most or all of the SDSS sub-exposures
were taken during phases in which the irradiated hemisphere was not facing the observer.

To verify the irradiated companion hypothesis, we fitted all the MMT spectra with a two-component model consisting of the best-fit
WD atmosphere model from the Koester grid (absorption component), and a set of six equal-strength Gaussian lines in the Balmer
series, from H$\gamma$ to H10 (emission component).
For these fits, we scanned over possible RVs $v_{a}$ and $v_{e}$ for the absorption and emission components, respectively. For 
each pair of RV values we fitted for the emission to absorption flux ratio, $F_{e}/F_{a}$ (forced to be positive), as well as an overall 
normalisation, described as a third-order polynomial in wavelength.  We then determined the best-fit values of the RVs and their 
uncertainties by a parabolic fit to the $\chi^2$ surface around the best-fit location, taking $F_e/F_a$ and its uncertainty from 
the fit to the nearest grid point.  For all spectra, we found good fits, with reduced $\chi^2$ values around~0.8.
The resulting values are listed in Table \ref{tab:RVs}, where we have corrected all times and radial velocities to the 
barycenter of the Solar System. 

In seven of the 19 spectra (MMT 5, 6, 7, 8, 12, 13, and 18; used to construct the spectrum in Fig.~\ref{MMTspect}), our procedure
did not detect any line emission: at the best-fit $v_a$, the addition of an emission component made for a worse fit at any value
of~$v_e$. Since random chance should lead to some non-zero contributions from the emission component even in spectra that can be
described by the absorption component alone, we inspected these fits more closely. We concluded that the absence of spurious
detections of an emission component even at unreasonable values of~$v_e$ results from our best-fit WD model slightly
underpredicting the depths of the line cores (see Fig.~\ref{MMTspect}), so that the best-fit emission component would have a
negative flux, which our fit routine does not allow. This is a minor effect that we did not try to correct for, but we note that it
implies that our $F_e/F_a$ flux ratios are slightly biased towards low values.

For the spectra with clear emission (MMT 1, 2, 3, 4, 9, 10, 11, 14, 15, 16, and 17), we find that the strength of the emission
lines varies quite a bit, with the lines appearing stronger at superior conjunction of the companion (see Figure \ref{Spectra} for
an ilustration of this variation in the case of H$\gamma$ - other lines show similar behaviour). The overlap between the cores of
the emission and absorption lines makes the determination of the component RVs somewhat challenging, which results in measurements
that are more noisy than would be expected in a single-lined system with spectra of similar quality.

\begin{table*}
  \centering
  \begin{minipage}{140mm}
    \centering
    \caption{Spectroscopic RVs for \wdth\ from the MMT spectra}
    \begin{tabular}{@{\extracolsep{\fill}}lrrrr}
      \hline
      Spectrum & MJD (bar.) & $v_{a} (\mathrm{km \, s^{-1}})$ & $v_{e} (\mathrm{km \, s^{-1}})$ & $F_{e}/F_{a}$ \\
      \hline
      MMT  0  & 55275.3824  &  $ -61.4 \pm   18.1$ &  $  69.2 \pm   10.1$ &  $1.24 \pm 0.14$ \\
      MMT  1  & 55275.3868  &  $ -51.7 \pm   21.7$ &  $   7.4 \pm   12.1$ &  $1.33 \pm 0.16$ \\
      MMT  2  & 55275.3925  &  $   3.8 \pm   14.7$ &  $ -64.9 \pm    7.7$ &  $1.52 \pm 0.11$ \\
      MMT  3  & 55275.4012  &  $  44.2 \pm   12.7$ &  $-188.7 \pm    7.3$ &  $1.25 \pm 0.11$ \\
      MMT  4  & 55275.4083  &  $  39.1 \pm   12.1$ &  $-243.9 \pm   10.8$ &  $0.74 \pm 0.10$ \\
      MMT  5  & 55275.4447  &  $ -39.7 \pm   15.9$ &  ...  &  $ 0.00 \pm  0.15$ \\
      MMT  6  & 55275.4518  &  $ -42.6 \pm   13.7$ &  ...  &  $ 0.00 \pm  0.13$ \\
      MMT  7  & 55275.4604  &  $ -83.1 \pm   14.2$ &  ...  &  $ 0.00 \pm  0.13$ \\
      MMT  8  & 55275.4675  &  $ -95.6 \pm   14.4$ &  ...  &  $ 0.00 \pm  0.14$ \\
      MMT  9  & 55275.4747  &  $ -64.6 \pm   15.3$ &  $ 187.1 \pm   28.1$ &  $0.25 \pm 0.13$ \\
      MMT 10  & 55277.3577  &  $  39.6 \pm   14.9$ &  $-299.0 \pm   15.4$ &  $0.55 \pm 0.14$ \\
      MMT 11  & 55277.3642  &  $ -18.9 \pm   20.5$ &  $-314.4 \pm   25.6$ &  $0.30 \pm 0.18$ \\
      MMT 12  & 55277.3718  &  $   7.2 \pm   15.6$ &  ...  &  $ 0.00 \pm  0.16$ \\
      MMT 13  & 55277.3783  &  $  15.8 \pm   16.2$ &  ...  &  $ 0.00 \pm  0.17$ \\
      MMT 14  & 55277.4556  &  $  53.5 \pm   17.9$ &  $-115.3 \pm   10.0$ &  $1.27 \pm 0.15$ \\
      MMT 15  & 55277.4620  &  $  33.4 \pm   18.2$ &  $-214.1 \pm   13.4$ &  $1.08 \pm 0.17$ \\
      MMT 16  & 55277.4697  &  $  69.6 \pm   24.7$ &  $-265.8 \pm   54.1$ &  $0.17 \pm 0.21$ \\
      MMT 17  & 55277.4762  &  $  45.6 \pm   19.5$ &  $-320.8 \pm   39.4$ &  $0.21 \pm 0.17$ \\
      MMT 18  & 55277.4826  &  $  13.8 \pm   14.1$ &  ... &  $ 0.00 \pm  0.14$ \\
      \hline
    \end{tabular}
    \label{tab:RVs}
  \end{minipage}
\end{table*}

\begin{figure}
  \includegraphics[width=0.32\textwidth,angle=90]{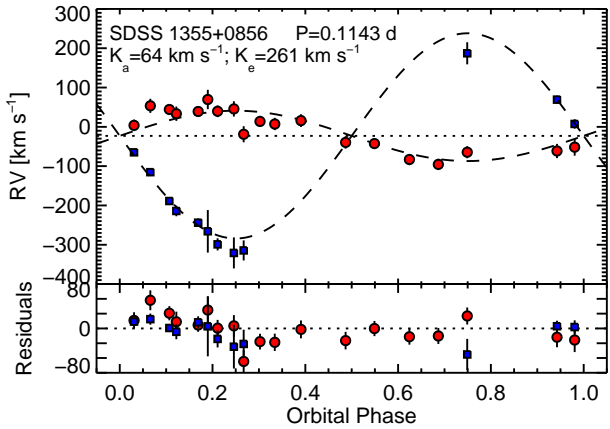}
  \caption{Orbital solution for \wdth. The red circles represent the RVs of the absorption component (WD), and the blue squares
    represent the RVs of the emission component in the spectra with $F_{e}/F_{a} > 0$. The value of $\gamma_{a}$ ($-23$ \kms) has
    been marked with a horiozontal dotted line.}
  \label{fig:Orbit}
\end{figure}

\begin{figure}
  \includegraphics[width=0.32\textwidth,angle=90]{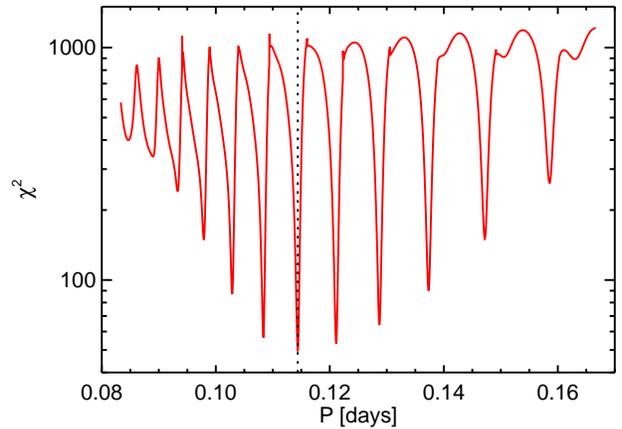}
  \caption{Periodogramme for the orbital solution of \wdth, with indication of the best-fit period ($0.11438$ days).}
  \label{fig:Pgam}
\end{figure}

We used the RVs to derive orbital solutions of the form $v = \gamma + K \sin{2 \pi (t-T_{0})/P}$ for each component. To get the
best possible constraints on the binary period, we fitted the emission and absorption RVs jointly, assuming that they have the
same period and anomaly, but are in antiphase. In our procedure, we scan a grid in period, at each point fitting for $T_0$, the
semi-amplitude of the absorption component $K_a$, its systemic velocity $\gamma_a$, the semi-amplitude of the emission component
$K_e$, and the difference in systemic velocity $\Delta \gamma = \gamma_e-\gamma_a$.  Our best-fit orbital solution, shown in
Figure~\ref{fig:Orbit}, has a reduced $\chi^2$ of 1.97 (for 25 d.o.f.), i.e. it is formally not acceptable. This is likely due to
inaccuracies of the spectral model. Since the residuals had no obvious phase dependencies, we compensated for the poor quality of
the fit by increasing all errors by $\sqrt{1.97}$. We find a period of $0.11438 \pm 0.00006$~d and epoch of mean longitude $T_{0}
= 55275.9613 \pm 0.0012$~MJD (barycentric).  The small error bar on the period results from fitting absorption and emission jointly,
as can be seen from the narrow $\chi^2$ minima in the periodogramme (see Figure \ref{fig:Pgam}).  However, we can only
exclude the alias at 0.12112~d at the 2$\sigma$ level (the two fits differ by $\Delta\chi^2=3.8$).  For the best-fit period, we
infer $K_{a}=64 \pm 8$ \kms\ and $K_{e}=-261 \pm 13$ \kms\ for the absorption and emission components, respectively.  The systemic
velocity for the absorption component is $\gamma_{a} = -23 \pm 5$ \kms, and that for the emission component identical within the
uncertainties, $\Delta \gamma = -4 \pm 18$ \kms.

\begin{figure}
  \includegraphics[width=0.32\textwidth,angle=90]{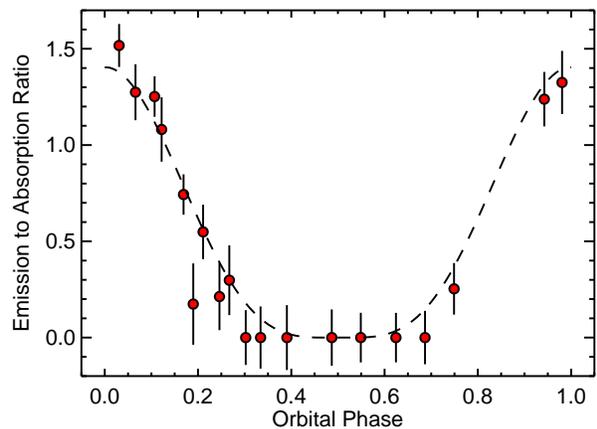}
  \caption{Behaviour of the $F_{e}/F_{a}$ ratio in the MMT spectra of \wdth\ as a function of orbital phase. The dashed line
    represents the model light curve discussed in Section~\ref{sec:Disc}.}
  \label{fig:EmToAb}
\end{figure}

This orbital solution nicely confirms the irradiated companion hypothesis. The spectra without detected line emission are only
found between phases 0.25 and 0.75, in the part of the orbit where the RV of the absorption component is decreasing and the
irradiated side of the companion is partially or totally occulted by the rest of the star. Outside of this phase range, a larger 
fraction of the irradiated side of the companion is in view, and the line emission can be easily detected. This scenario makes
a specific prediction about the strength of the line emission, which should peak around phase 0, fall down gradually as the system
approaches quadrature (phase 0.25), stay low until quadrature is reached again (phase 0.75), and then increase again. This is
indeed the behaviour of the $F_{e}/F_{a}$ ratio in the MMT spectra of \wdth\ when it is folded through the best-fit orbital
solution (see Figure~\ref{fig:EmToAb}). The numerical value of this ratio is arbitrary, because it depends on the normalization of
the emission and absorption models that we used to fit the spectra,
but
qualitatively its orbital evolution matches that of the equivalent width in the line emission of GD 448
\citep{marsh1996MNRAS.278..565M:GD448,maxted1998MNRAS.300.1225M:GD448}. For GD 448, the equivalent widths of the emission lines do
not go to zero between phases 0.25 and 0.75, but that is expected given the relatively low orbital inclination of the system
\citep[$29.3^\circ\pm0.7^\circ$][]{maxted1998MNRAS.300.1225M:GD448}. The absence of detectable line emission during these phases
and the larger RVs imply a higher inlcination for \wdth.

\section{Discussion}
\label{sec:Disc}

For the values of $T_{\rm eff}$ and $\log{g}$ determined from the co-added SDSS spectrum, the models of \cite{Panei2007} for He
WDs give a mass of $M_{1}=0.46\pm0.01$ \msun, and those of \cite{Holberg2006} for C/O WDs give a slightly lower value of
$M_{1}=0.43 \pm 0.01$~\msun. Since these values are both below the $\sim\!0.48$~\msun\ threshold for C ignition in a stellar core
\citep{sweigart1978}, we conclude that the primary in the \wdth\ system is most likely a hot He-core WD, and we adopt the mass
given by the He-core models \citep[however, see][for the possibility that it might be a hybrid CO-He WD]{PradaMoroni2009}.

Even without considering the RVs measured for the line emission component, the mass of the companion is tightly constrained by the
broadband SED of the system and the mass function of the WD primary derived from the absorption RVs. A quick comparison between
the SED shown in Figure~\ref{fig:SED} and the theoretical models of \cite{Baraffe1998} rules out companions more massive than
$\sim\!0.1$~\msun, which would lead to noticeable excess flux in the IR, especially in the $J$, $H$, and $K$ bands. More formally,
we can set an upper limit on $M_{2}$ by requiring that the $K$-band flux predicted by the stellar models does not exceed the
observed flux from UKIDSS by more than a factor 10. This generous tolerance should accomodate both the issues identified with
theoretical low-mass stellar models \citep{Casagrande2008,Kraus2011} and the uncertainties associated with the UKIDSS photometry
\citep{Leggett2011}. By interpolating on the mass grid of the \cite{Baraffe1998} models, we determine a conservative upper limit
of $0.097$ \msun\ for $M_{2}$. The mass function for the primary also puts constraints on the companion mass, requiring
$M_{2}\sin{i} = 0.098\pm0.016$ \msun, where the error bar reflects the observational uncertainty in the values of $K_{1}$ and
$M_{1}$.  From these limits, we can conclude that the nondegenerate companion of \wdth\ is a low-mass main sequence star with a
mass between $0.083$ and $0.097$ \msun, above the upper mass limit for brown dwarfs. The star should have a spectral type between
M5 and M8 \citep{Baraffe1996}, and be fully convective \citep{Reiners2009}. Our faliure to detect a significant offset between the
RV curve of the WD and that of the line emission from the companion ($\Delta \gamma = -4 \pm 18$ \kms\ in our fit) does not allow
us to put further constraints on the component masses using gravitational redshift, although this should be possible with data of
higher quality \citep[see discussion in][their Section 3.4, for the case of GD 448]{maxted1998MNRAS.300.1225M:GD448}.

The RV curve that we have measured for the line emission provides an independent confirmation of our estimate for the companion
mass. To interpret these RV measurements we need to account for the fact that the line emission does not originate from the entire
surface of the companion, so the center of light for the line emission does not correspond to the center of mass for the star, but
is shifted towards the WD. This means that the measured semiamplitude of the emission component, $K_{e}$, is only a lower limit to
the true semiamplitude of the RV curve of the companion ($K_{2}$). This effect has been observed in other systems with hot WD
primaries like GD 448 \citep{marsh1996MNRAS.278..565M:GD448} and SDSS J212531.92$-$010745.9 \citep{Nagel2006,Schuh2009}, and is
quite common in low mass X-ray binaries with NS primaries surrounded by an accretion disk \citep[see][and references
therein]{MunozDarias2005}. 

\begin{figure}
  \includegraphics[width=0.32\textwidth,angle=90]{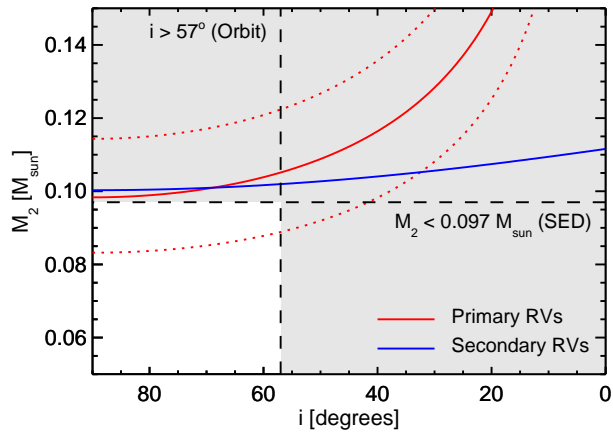}
  \caption{Observational constraints on $M_{2}$ and $i$ for \wdth. The horizontal dashed line is the upper limit on $M_{2}$
    derived from the SED (Figure \ref{fig:SED}). The vertical dashed line is the lower limit on $i$ imposed by the measured value
    of $K_{e}$. The shaded regions above and to the right of these lines are forbidden by the data. The solid red line represents
    the constraint on $M_{2}\sin{i}$ from the mass function of the primary, with the dotted red lines indicating standard
    1$\sigma$ uncertainties. The solid blue line represents the constraint from the measured value of $K_{e}$ and the light curve
    model shown in Figure \ref{fig:EmToAb}. The (rather large) uncertainty on this constraint is not shown.}
  \label{fig:M2sini}
\end{figure}

For a detached binary without an accretion disk, $K_{e}/K_{2} = 1 - f(R_{2}/a_{2})$, where $a_{2}$ is the orbital radius of the
companion around the barycentre of the system, $R_{2}$ is the companion radius, and $f$ is a dimensionless factor that depends on
the extent of the irradiated area on the surface of the companion and on the orbital inclination of the binary
\citep{VanKerkwijk2011}. In practice, the orbital inclination has a very small effect on the value of $f$ for $i \gtrsim
50^{\circ}$ \citep{MunozDarias2005,VanKerkwijk2011}. We can estimate $f$ by assuming the emission from the irradiated hemisphere
is proportional to the incident flux from the primary.  For an irradiating source at a large distance, the integral has an
analytic solution at quadrature, and $f=3\pi/16=0.59$, with higher values of $f$ If the stars are closer together.  For \wdth, we
expect $R_2\simeq0.1\,R_\odot$ (for a very low-mass M dwarf, \citealt{Baraffe1998}), $a\simeq0.8\,R_\odot$ (for a total mass of
$\sim\!0.59\,M_\odot$), and thus $R_2/a=1/8$.  For that value, numerical integration yields $f=0.66$ at quadrature, with a range
of $\pm0.02$ for $1/10<R_2/a<1/6$.  From Figure~\ref{fig:EmToAb}, one sees that such a model reproduces the flux ratio between the
emission and absorption components in our spectral fits quite well.  Trying different inclinations, we find that the absence of
emission at phases 0.3--0.7 requires an inclination $i\ga52^\circ$ (for smaller inclination, the flux at phase 0.3 would exceed
20\% of the maximum flux).
After some algebra, we can write an expression for $M_{2}$ which depends strongly on measured parameters
and weakly on $R_{2}$ and $\sin{i}$:
\begin{equation}
M_{2}=\frac{M_{1}K_{1}}{K_{e} + 2 \pi f \sin{i} (R_{2}/P)} .
\end{equation}
For the nominal values of $M_{1}$, $f$, $K_{1}$, and $K_{e}$, and $i=90^{\circ}$, this yields $M_{2}=0.10$ \msun, which agrees quite
nicely with our constraints from the mass function of the primary and the SED. Unfortunately, the error bars on $K_{1}$ and
$K_{e}$ are large enough that this method cannot refine our $M_{2}$ estimate any further, just confirm it. In any case, the measured
value of $K_{e}$ does put an upper limit on the orbital inclination,
\begin{equation}
\sin{i} \geq \frac{K_{1}+K_{e}}{(2 \pi G (M_{1}+M_{2})/P)^{1/3}},
\end{equation}
which translates to $i \geq 65 \pm 7 ^{\circ}$ (or a lower limit of 57$^{\circ}$), with the uncertainty on the inclination angle
is again dominated by the errors on $K_{1}$ and $K_{e}$.  The limit is in line with our expectations from the mass constraints and
the emission line lightcurve. We summarize our constraints on $M_{2}$ and $i$ in Figure \ref{fig:M2sini}.

The picture that emerges from our analysis of \wdth\ is that of a binary similar to GD 448
\citep{marsh1996MNRAS.278..565M:GD448,maxted1998MNRAS.300.1225M:GD448}. The two systems have almost identical component masses and
period, but \wdth\ has a somewhat higher inclination, leading to higher RVs and no detectable line emission between phases 0.25
and 0.75. The cooling age of the WD in \wdth\ is $\sim 25$ Myr \citep{Panei2007}, about half the value for GD 448. This cooling
time is much shorter than both the thermal timescale of the companion ($\sim 5$ Gyr) and the timescale for orbital decay due to
gravitational wave emission ($\sim 2$ Gyr). The system was probably born in the period gap of CVs, emerging from the CE phase with
essentially the same period we can observe today. The possibility that \wdth\ is instead crossing the period gap after a CV phase
is unlikely, because (a) quiescent CVs usually have much cooler WDs \citep{Townsley2002} and (b) it is hard to imagine how mass
transfer could have happened at a longer period with such a low mass companion. Since magnetic braking is inefficient for systems
with fully convective secondaries \citep{Schreiber2003,Zorotovic2011}, the future evolution of \wdth\ will be dominated by
gravitational wave radiation, but the system will still become semi-detached in less than a Hubble time, much like GD 448
\citep[see Table 2 in][and accompanying discussion]{Schreiber2003}. The low value of $q$ ($0.21$) implies that mass transfer will
be stable \citep{Paczynski1971,King1996}. At this point, \wdth\ will become one of the rare CVs with likely He-core WD primaries,
possibly leading to a few exceptionally long classical nova events \citep{Shen2009}. The period and component masses of the system
are such that it will contribute to future demographic constraints of CE evolution \citep[see discussion in][]{Zorotovic2011a},
but we leave that analysis for further work.




\section*{Acknowledgments}

We wish to thank Detlev Koester for making his WD atmosphere models available to us. Balmer/Lyman lines in the Koester WD models
were calculated with the modified Stark broadening profiles of \cite{tremblay2009ApJ...696.1755T:Stark_Broadening}, kindly made
available by the authors. We are indebted to Scott Kleinman, who made his WD catalogue available to us in advance of
publication. We are grateful to Eduardo Bravo, Avishay Gal-Yam, Shri Kulkarni, Dan Maoz, Thomas Matheson, Gijs Nelemans, Benny
Trakhtenbrot, and Sharon Xuesong Wang for discussions.

Funding for the SDSS and SDSS-II has been provided by the Alfred P. Sloan Foundation, the Participating Institutions,
the National Science Foundation, the U.S. Department of Energy, the National Aeronautics and Space Administration, the
Japanese Monbukagakusho, the Max Planck Society, and the Higher Education Funding Council for England. The SDSS Web Site
is \url{http://www.sdss.org/}. The SDSS is managed by the Astrophysical Research Consortium for the Participating
Institutions. The Participating Institutions are the American Museum of Natural History, Astrophysical Institute
Potsdam, University of Basel, University of Cambridge, Case Western Reserve University, University of Chicago, Drexel
University, Fermilab, the Institute for Advanced Study, the Japan Participation Group, Johns Hopkins University, the
Joint Institute for Nuclear Astrophysics, the Kavli Institute for Particle Astrophysics and Cosmology, the Korean
Scientist Group, the Chinese Academy of Sciences (LAMOST), Los Alamos National Laboratory, the Max-Planck-Institute for
Astronomy (MPIA), the Max-Planck-Institute for Astrophysics (MPA), New Mexico State University, Ohio State University,
University of Pittsburgh, University of Portsmouth, Princeton University, the United States Naval Observatory, and the
University of Washington.

{\it Facilities:} MMT (Blue Channel Spectrograph); ARC 3.5m Telescope (Dual Imaging Spectrograph).


\end{document}